\documentclass[aps,prl,superscriptaddress,twocolumn,showpacs]{revtex4}
\usepackage{graphicx}
\begin{document}

\title{Coarsening of granular clusters: two types of scaling behaviors}

\author{M.V. Sapozhnikov}
\affiliation{Argonne National Laboratory, 9700 South  Cass Avenue,
Argonne, IL 60439, USA}
\affiliation{ Institute for Physics of
Microstructures, Russian Academy of Sciences, GSP-105, Nizhny
Novgorod, 603600, Russia}
\author{I.S. Aranson}
\affiliation{Argonne National Laboratory, 9700 South  Cass Avenue,
Argonne, IL 60439, USA}
\author{J.S. Olafsen}
\affiliation{Department of Physics \& Astronomy, University of
Kansas, Lawrence, KS 66045, USA}

\date{\today}

\begin{abstract}
We report on an experimental study of small cluster dynamics during
the coarsening process in driven granular submonolayers of $120\mu m$
bronze particles. The techniques of electrostatic and vertical
mechanical vibration were employed to excite the granular gas. We
measure the scaling exponent for the evaporation of small clusters
during coarsening. It was found that the surface area of small
clusters $S$ vs time $t$ behaves as $S \sim (t_0-t)^{2/3}$ for
lower frequencies and $S\sim (t_0-t)$  for higher frequencies. We
argue that the change in the scaling  exponent is related to the
transition from three dimensional (3D) to two dimensional (2D)
character of motion  in the granular gas.
\end{abstract}

\pacs{45.70.Mg,45.70.Qj}

\maketitle One of the most fascinating phenomena observed in
freely-cooling  and driven granular gases is spontaneous
separation into dilute and dense regions \cite{jaeger,kadanoff}.
This phenomenon
stimulated great interest among theorists and experimentalists
\cite{intro,losert99,kudrolli97,olafsen98}.
The clustering phenomenon appears to not be
limited to just the traditional granular systems with short-range
hard-core collisions. It was shown that some granular  systems
with competing long-range electromagnetic interactions and
short-range collisions exhibit somewhat similar behavior
\cite{aranson00}.  In the case of mechanically vibrated granular
matter where the energy is injected in the system through local
collision with the bottom plate, clustering takes place due to the
increase of the dissipation rate with the increase of the density
of granular gas \cite{kudrolli97,olafsen98}. In the case of
electrostatically driven granular matter there is an additional
effect of electrostatic screening which decreases the external
electrostatic force acting on particles inside dense clusters
\cite{aranson00}.

The phase diagram for these two seemingly different
systems  appears to be rather independent on the specifics of
driving \cite{olafsen98,losert99,aranson00}. Particles are
immobile for small values of the driving force (amplitude of
vibration acceleration in the case of mechanical system and the
amplitude of electric field for the electrical system). When the
driving force exceeds the first critical value, isolated particles
begin to move. If the driving force value is increased above a
second threshold value, the granular medium forms a nearly uniform
granular gas. Between these values the uniform
gas phase is unstable  and a fraction of particles can form
immobile clusters. The clusters exhibit coarsening  dynamics of
Ostwald-ripening type: small clusters disappear  and  large
clusters grow with time.

Here  we report experimental study of small cluster
dynamics during the coarsening process  in driven granular
submonolayers.  We systematically compare two techniques of
energizing the granular gas: mechanical and electrostatic. We
measure the scaling exponent for the evaporation of small clusters
during coarsening. We have found that in both systems the surface
area of small clusters $S$ vs time $t$ behaves as $S \sim
(t_0-t)^{2/3}$ for lower frequencies and $S\sim (t_0-t)$ for
higher frequencies. This transition in scaling behavior coincides
with the transition from 3D dynamics of the granular gas to 2D
dynamics as the driving frequency is increased for a fixed
acceleration.

The experimental setup for mechanical shaking is similar to that
in \cite{umbanhovar96,olafsen98,aranson99,losert99}. We performed
experiments in a rigid circular container of $13.5$ cm diameter
vibrated vertically by an electromagnetic shaker. The spherical
bronze particles ($120 - 150 \mu m$) constituted approximately 2/3
of a monolayer coverage on the optically-flat horizontal plate
(single-crystal Silicon wafer).
The applied frequency varied in the range of 20-100 Hz. Most of
the measurements were performed under atmospheric pressure in air.
Some results were verified in vacuum.

To excite a granular media electrostatically, particles were placed
between plates of a large capacitor and energized
either by DC or AC electric fields \cite{aranson00}. We used
$27\times 27$ cm transparent capacitor plates (glass coated by
indium doped tin oxide) and the plate spacing was
1.5 mm.  The particles used for these series of
experiments are the same as for the mechanical shaking
measurements and constituted approximately 1/3 of a monolayer
coverage on the bottom plate. The
applied field was varied in the range  $0-7$ kV/cm and its
frequency was varied from 0 to 120 Hz. The experiments were
performed in an atmosphere of dry nitrogen to reduce adhesion of
particles on the plate due to humidity in air \cite{howell}.  The
CCD camera was suspended  above both cells. We analyzed the total
number of clusters, evolution of individual clusters, mean cluster
sizes, etc.

In the
electro-cell, energy injection works as follows
\cite{aranson00}. Particles in the contact with bottom plate
acquire an electric charge and the electric force exerted on
the particle is $F_0=1.36 R^2 E^2$, $R$ is the radius of the
particle, $E$ is the external field. When $F_0$ exceeds the
gravity, the particle moves upwards, recharges upon colliding with
the upper plate and falls down. In the case of DC fields the
process repeats in a cyclical manner. By applying AC field one may
turn the particle back before it reaches upper plate. Thus, increasing
the frequency of the field on controls the extent of vertical
motion of the particles. Clustering of granular gas with
consequent ripening occurs if $E_1<E<E_2$ ($E_1$ and $E_2$ is the
first and second threshold values). The clustering comes from two
effects: increased dissipation of the kinetic energy in the
regions with high granular media density and electrostatic
screening (two close particles acquire smaller charge then two
well-separated ones). In the mechanical cell, vibration amplitude
plays the role of electric field.

The primary objective of our studies was to observe the scaling behavior
of individual clusters as function of time. In certain cases one can
relate the scaling exponent governing evaporation of small clusters
with  global scaling properties of coarsening process due to
conservation laws \cite{meerson,hannon}. For example, if  the area
of small cluster $S$ in 2D  system  evolves as $S\sim
(t_0-t)^\beta$, where $t_0$ is the instant of the cluster's
disappearance and $\beta$ is scaling exponent, and the total
number of clusters $N$ evolves as $N\sim 1/t^\alpha$, for
interface-controlled coarsening regime the exponents $\alpha$ and
$\beta$ are related, $\alpha=\beta$ \cite{meerson}. In many cases
the studies of individual cluster dynamics are more reliable than
direct studies of global scaling exponents which are often masked
by the finite size effects.

\begin{figure}[ptb]
\includegraphics*[width=8cm]{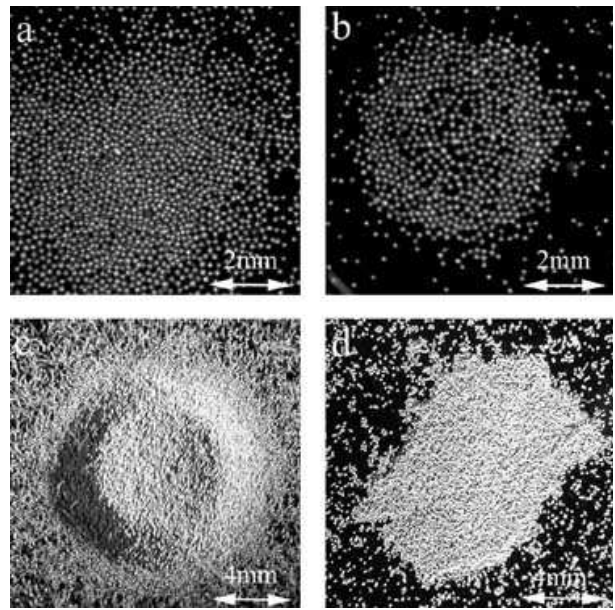}
\caption{Shapes of granular clusters. Upper row shows clusters in
electrostatically driven system. a) $f=20$Hz; b) $f=120$Hz. Bottom
row shows clusters in mechanically vibrated system. c) $f=25$Hz;
d) $f=95$Hz. Left column: heap-like clusters corresponding to 3D
behavior. Right column: flat monolayer clusters corresponding to
2D regime. \label{clusters}}
\end{figure}

The experiments in the mechanical cell  were performed in the
following manner. A uniform gas-like phase was prepared by setting
the dimensionless acceleration level $\Gamma=4 \pi^2 f^2 A_0/g$
($f$ is vibration frequency, $A_0$ is the amplitude of driving
$A=A_0 \sin (2\pi ft)$, $g$ is gravity acceleration) approximately
at $\Gamma=2.28 - 2.72g$.  The acceleration was then suddenly
dropped  to the value $1.00 - 1.05$ leading to spontaneous
nucleation of immobile clusters from the gas phase
(Fig.\ref{clusters} c,d).

\begin{figure}[ptb]
\includegraphics*[width=8cm]{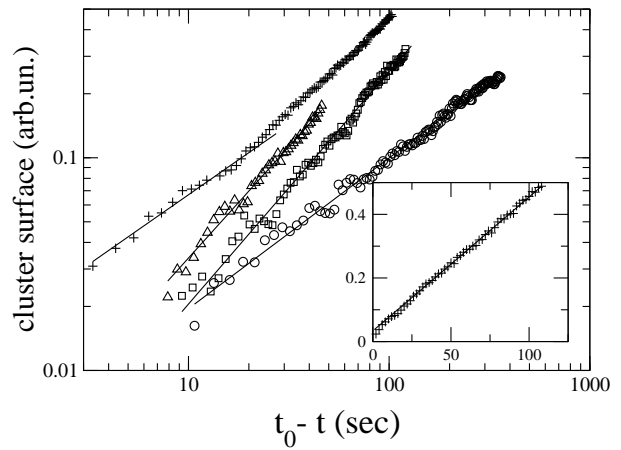}
\caption{Cluster area vs. time for mechanical system.
Circles: $f=25$Hz, $\beta=0.7$. Crosses: $f=35$Hz.  $\beta=1.03$ at
the beginning of evaporation (see inset)  and $\beta=0.66$ at the end of
collapse. Triangles: $f=55$Hz, $\beta=1.04$. Squares: $f=85$Hz,
$\beta=0.7$. Inset: $S$ vs $t$ for $f=35$Hz in linear scale
\label{shaking} }
\end{figure}

The area of a selected cluster $S$ was measured as a function of
time. If the cluster is not the biggest one in the
system, eventually  it begins to shrink. We find that the area $S$
evolves according to  a power law with an exponent that depends on
the vibration frequency. For frequencies less than 35Hz, $S \sim
(t_0-t)^\beta$ with $\beta=0.63 - 0.7$,  see Fig.\ref{shaking}.
For the frequencies higher than 55Hz, $S \sim (t_0-t)^\beta$ with
$\beta\approx 1 $. A crossover-type  behavior observed for an
intermediate range of frequencies. In this case the area
evolves  as follows: while the cluster is above a certain size,
its area decreases linearly with time. However, when the cluster
decreases below a certain size, its area decreases quicker, as
$S\sim (t_0-t)^\beta$ with $\beta\approx 2/3$.

\begin{figure}[ptb]
\includegraphics*[width=8cm]{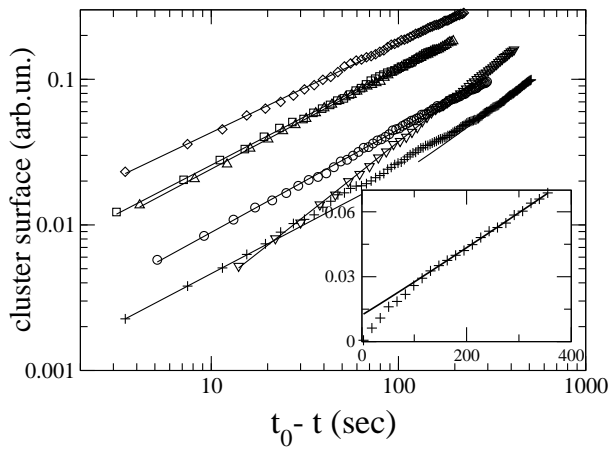}
\caption{Cluster area vs. time for electro-cell. Circles: $f=0$,
$\beta=0.7$. Squares: $f=10$Hz, $\beta=0.67$. Diamonds: $f=20$Hz,
$\beta=0.63$. Up triangles: $f=50$Hz, $\beta=0.68$. Crosses:
$f=75$Hz, $\beta=1.04$ at the initial stage of evaporation (see inset)
and $\beta=0.67$ at the end. Down triangles: $f=100$Hz,
$\beta=1.02$.  Inset: $S$ vs $t$ for $f=75$Hz in linear scale \label{electro} }
\end{figure}

Qualitatively similar  behavior is observed in an
electrostatically driven system (Fig.\ref{electro}).
In the
electro-cell, the uniform gas was prepared by applying an external
field of $7$ kV/cm before "quenching" the gas below $E_2$.
For the
case of a DC electric field applied, as well as for an AC
field with the frequency less than 60Hz, we find that the scaling
exponent  $\beta \approx 2/3$ (actually the range is $0.63 -
0.7$). Accordingly, the cluster area shrinks linearly in time for
frequencies above 90Hz. In the range from 60 to 90Hz, there is a
crossover behavior observed with these two exponents.

The apparent frequency dependence of the scaling exponents $\beta$
for cluster evaporation can be related to the transition from two
to three dimensional behavior in the granular gas.
For the case of DC or  low  frequency AC driving, the
particles have enough time to reach the upper plate and the
granular gas exhibits 3D behavior. Thus,  gas particles can fly
above the cluster and the exchange of particles between gas and
solid phases occurs at entire surface of the cluster.
Consequently, the  clusters  contain more than a monolayer of the
particles (Fig. \ref{clusters}a). Measurements show that in the
electrostatic cell with a gap equal to 1.5 mm, cluster thicknesses
equal to 4 monolayers are observed in the central part of the
cluster. The thickness is limited due to an increase of the
electric field above the cluster as the gap between its surface
and upper plate shrinks. When the field value at the cluster
surface reaches the second threshold value $E_2$, particles
cannot remain immobile, even in the middle of the cluster, and
will rejoin the gas.  As the extent of  vertical particle motion
becomes smaller than the particle  diameter  (for high
frequencies), the granular gas becomes two dimensional. In this
case, the exchange of particles between gas and solid
takes place only at the outer boundary  and
clusters have a monolayer thickness  (Fig \ref{clusters}b).

Similar transition occurs in
mechanically vibrated cell.  If we keep the value of the
dimensionless acceleration $\Gamma$ fixed, lowering of the frequency
implies  shaking at higher amplitudes, and,
consequently, increase of the extent of vertical motion.
Thus, at lower frequencies the vertical particle jumps will
eventually exceed the particles size $d$, i.e. the motion is
effectively  more 3D than it is at higher frequencies.  The
dimensional transition in the collisions dissipation rate
has been
previously observed in Ref.
\cite{olafsen99}.

The crossover frequency can be determined from the comparison of
the height of particle jumps $\xi$ and the particle size $d$. In
mechanical system the height  may be roughly estimated to be equal
to vibration amplitude $A_0$, i.e. $\xi=g/4 \pi^2 f^2 $  since
clusterization occurs when $\Gamma\approx 1$.  From the condition
$\xi \approx d$ one find the crossover frequency  $1/2\pi
\sqrt{g/d}=40$Hz for $d=150\mu$m, which is in a good agreement
with experiment. The value of $\xi$ decreases as the
frequency increases and when it becomes smaller than a particle diameter,
the granular gas exhibits 2D dynamics. In this case clusters are
flat (Fig. \ref{clusters}d). As the frequency decreases, the
magnitude of vertical jumps increases and the clusters assume
heap-like form   (Fig. \ref{clusters}c). Their evaporation
will reflect features of   3D gas dynamics as is the case of
electrostatic excitation. The saturation of the cluster occurs
when the cluster hight achieves the magnitude of the vertical
jumps of isolated particles. It is necessary to mention that this
heap-like form of the clusters is not necessarily related  with
the interaction between particles and trapped air flow through the
bulk of the medium as it was in Refs. \cite{heap}. Clusters keep
heap-like form even in the  evacuated cell (pressure is 5mTorr).


According to the model presented in the \cite{aranson02} when the
cluster size becomes small (i.e. near the  moment of evaporation)
the normal velocity component of the interface between solid and
gas phase for 2D dynamics is governed by equation $\dot{R}\approx -K/2R$.
Here $R$ is the local radius of the interface, $K$ is surface
tension coefficient which can be expressed through the diffusivity
of  particles. Thus, for the cluster area $S= \pi R^2$ one derives
$S =\pi K(t_0-t)$, i.e $S$ decreases linearly during cluster
evaporation.

To match  the low-frequency scaling one assumes that
the evolution of small clusters is governed by equation
$\dot{R}\approx -\tilde K/R^2$. It yields  $R^3 \sim (t_0-t)$ and the
area occupied by the clusters scales as $S  \sim
(t_0-t)^{2/3}$. This result can be interpreted as a manifestation
of the decrease of the surface tension coefficient $K$ in
multilayered clusters. One expects  that the coefficient $K$ is
inversely proportional to the cluster's thickness $h$ (it requires
more time to evaporate a multilayer than a monolayer), i.e. $K
\sim h^{-1}$. Since for heap-like clusters the thickness $h$ is
proportional to the radius $R$, one obtains $\dot{R}\sim R^{-2}$.
This argument applies both to mechanical and electo-cells.

We would like to emphasize that the low-frequency
exponent $\beta=2/3$ appears to not be related to the
global scaling exponent $\alpha$  for the number of clusters $N$.
This is due to saturation of the height of large clusters: when a
cluster is large enough ($R \gg  h$) the dynamics of its
evaporation has a 2D character, as the gas/solid exchange occurs
mostly at the edge of the cluster. However,  when the cluster size
decreases to $R \sim h$, evaporation evaporation occurs on entire
surface of the cluster.

\begin{figure}[ptb]
\includegraphics*[width=8cm]{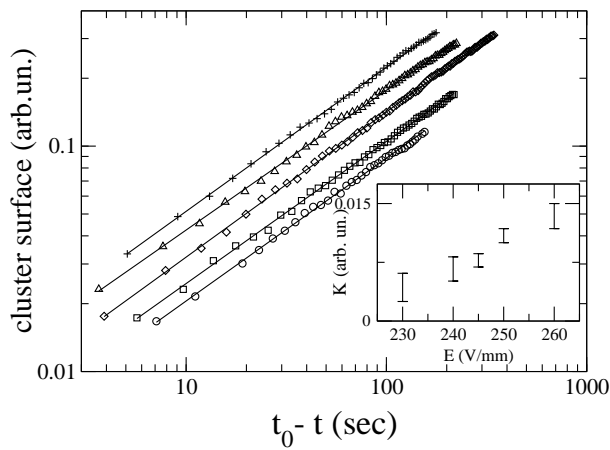}
\caption{Cluster area vs. time for electro-cell.
$f=20$Hz. Circles: electric field $E=230V/mm$. Squares:
$E=240V/mm$. Diamonds: $E=245V/mm$. Triangles: $E=250V/mm$.
Crosses: $E=260V/mm$. Inset: the dependence of cluster surface
tension on the applied electric field. \label{tension} }
\end{figure}

We find that the amplitude of driving mostly affects the pre-exponential factor
in the scaling law, i.e. the surface tension. Some selected results are shown
in Fig. \ref{tension}.
As one sees from the figure, the value of the surface tension
(and, therefore, diffusivity coefficient for the particles)
increases with the increase of the applied electric field. This
behavior can be attributed to the increase of the granular gas
temperature.
\begin{figure}[ptb]
\includegraphics*[width=8cm]{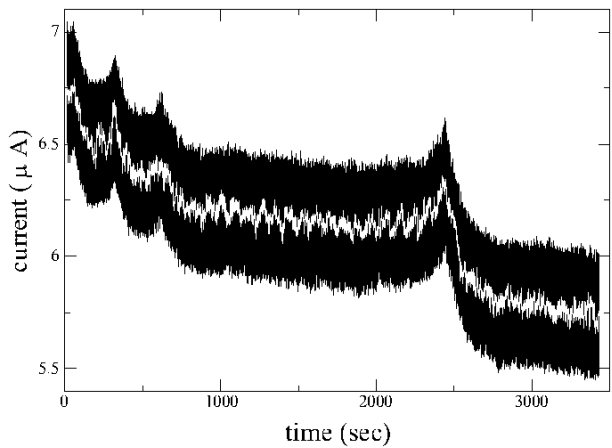}
\caption{Current vs. time during coarsening in DC electric field.
White line shows the signal averaged over 100 points. Current
peaks correspond to evaporation of clusters. The signal/noise ratio is better than 3.
\label{peaks} }
\end{figure}

One of the interesting predictions of the phenomenological model
\cite{aranson02}
is a fine structure in dependence of granular gas
density vs. time during the coarsening. A small cluster
shrinks so quickly at its evaporation point that the gas concentration
temporarily increases \cite{aranson95,meerson96,aranson02} resulting in
a peak feature. The electric current through the cell is carried by the
particles belonging to the granular gas phase and is, therefore,
proportional to the amount of gas. This leads to a transient
increase of the current through the system at the moment of small
cluster evaporation. These peaks of current were indeed observed
for the case of electrostatic cell with applied DC voltage (Fig.
\ref{peaks}).
Over
time, the current approaches a value corresponding to the
coexistence of one cluster and the granular gas. The observation
of the  peaks reinforces the validity of the continuum  model
described in \cite{aranson02}.

In conclusion, we studied experimentally the cluster dynamics of evaporation
in two driven granular systems, mechanical and  electrical. The transition
between 3D and 2D behaviors was found.
These results are  qualitatively similar for both systems.
For the cluster coarsening   process in a granular gas
excited by a DC electric field, the transient increase of
current through the system at the moment of a small cluster
evaporation was observed.
We are grateful to B. Meerson for fruitful discussions. This
research was supported by the US DOE, grant W-31-109-ENG-38.

\end{document}